\documentstyle[11pt,mrs2003,epsfig]{article} 
  
\newcommand\etal{\mbox{\textit{et al.}}} 
 
\newcommand\eg{{e.g.},\ } 
\def\ie{{i.e.},~}   
\def\4he{$^4$He}   
\def\3he{$^3$He}   
\def\7li{$^7$Li}   
\def\Yp{Y$_{\rm P}$~}   
\def\yd{$y_{\rm D}$~}   
\def\y3{$y_{3}$~}   
\def\y7{$y_{7}$~}   
\def\hii{H\thinspace{$\scriptstyle{\rm II}$}~}   
\def\hi{H\thinspace{$\scriptstyle{\rm I}$}~}   
\def\di{D\thinspace{$\scriptstyle{\rm I}$}~}   
   
\newcommand{\obh}{$\Omega_{\rm B} h^2\;$}

\newcommand{\omb}{$\Omega_{\rm B}\;$}

\newcommand{\be}{\begin{equation}}   
\newcommand{\ee}{\end{equation}}   
\newcommand{\Deln}{$\Delta N_\nu\;$}   
\def\Nnu{$N_{\nu}$~}   
\newcommand{\epm}{$e^{\pm}\;$}   
\def\be{\begin{equation}} 
\def\ee{\end{equation}}

\begin{document}  
\title{\begin{center} 
FORENSIC COSMOLOGY: \\  
PROBING BARYONS AND NEUTRINOS WITH BBN AND THE CBR 
\end{center}}  
 
\author{\begin{center}GARY STEIGMAN\end{center}} 
 
\affil{\begin{center}Departments of Physics and Astronomy,  
The Ohio State University\end{center}}

\begin{abstract}  
The primordial abundances of the light nuclides produced by Big Bang 
Nucleosynthesis (BBN) during the first 20 minutes in the evolution of 
the Universe are sensitive to the universal density of baryons and to 
the expansion rate of the early Universe.  For example, while deuterium  
is an excellent baryometer, helium-4 provides an accurate chronometer.   
Some 400 kyr later, when the cosmic background radiation (CBR) was  
freed from the grasp of the ionized plasma of protons and electrons,  
the spectrum of temperature fluctuations also depended on (among other 
parameters) the baryon density and the density in relativistic particles.  
The comparison between the constraints imposed by BBN and those from the  
CBR reveals a remarkably consistent picture of the Universe at two  
widely separated epochs in its evolution.  Combining these two probes  
leads to new estimates of the baryon density at present and tighter 
constraints on possible physics beyond the standard model of particle 
physics.  In this review the consistency between these complementary 
probes of the universal baryon density and the early-Universe expansion 
rate is exploited to provide new constraints on any asymmetry between 
relic neutrinos and anti-neutrinos (neutrino degeneracy). 
\end{abstract}  
  
\section{Introduction}\label{sec:intro} 
 
As the hot, dense, early Universe expanded and cooled, it briefly  
evolved through the epoch of big bang nucleosynthesis (BBN), leaving  
behind the first complex nuclei: deuterium, helium-3, helium-4, and  
lithium-7.  The abundances of these relic nuclides were fixed by the  
competition between the relative densities of nucleons (baryons) and  
photons (the baryon abundance) and the universal expansion rate.  In  
particular, primordial deuterium is an excellent baryometer, while  
the relic abundance of \4he provides an accurate chronometer.  Some  
400 thousand years later, when the cosmic background radiation (CBR)  
cooled sufficiently to allow neutral atoms to form, freeing the CBR  
from the embrace of the ionized plasma of protons and electrons, the 
spectrum of temperature fluctuations imprinted on the CBR recorded the 
baryon and radiation densities, along with the expansion rate of the 
Universe at that epoch.  As a result, the relic abundances of the 
light nuclides and the CBR temperature fluctuation spectrum provide 
invaluable, complementary windows on the early evolution of the Universe 
and sensitive probes of its particle content. 
  
In several, recent, related articles \cite{gs03}, the author has  
reviewed the current status of BBN and compared the BBN constraints  
on the baryon abundance and the early-Universe expansion rate with similar  
ones from the CBR (mainly from the WMAP data \cite{wmap} and analyses  
\cite{sperg}).  In a nutshell, the BBN-inferred baryon density based 
on the relic abundance of deuterium is in near-perfect agreement with 
that derived from the CBR. Although the value of the primordial 
abundance of \4he derived from observations of metal-poor (nearly 
primordial), extragalactic \hii regions is low compared to the standard 
BBN (SBBN) prediction, suggesting a slower-than-standard early-Universe 
expansion rate, the SBBN prediction is within $\sim 2\sigma$ of the data  
and is entirely consistent with the weaker constraint on the expansion  
rate some 400 thousand years later provided by the CBR. This consistency  
of the standard model of cosmology permits the exploration of more exotic 
alternatives, one of which -- a small but significant  asymmetry between  
neutrinos and antineutrinos (neutrino degeneracy) -- is explored here  
(see, also, Barger \etal~(2003a) \cite{barger03a} and references therein). 
 
To set the stage for the discussion of the cosmological constraints on 
neutrino degeneracy (see, \eg Kang \& Steigman 1992 \cite{ks}), we begin  
with a brief overview of SBBN and discuss the modifications to SBBN and  
the CBR in the presence of a non-standard, early-Universe expansion rate;  
for further details and references to related work, see Steigman (2003a,b,c)  
\cite{gs03}. 
 
\section{SBBN: An Overview}\label{sec:sbbn} 
  
Discussion of BBN can begin when the Universe is a few tenths of a second    
old and the temperature is a few MeV.  At such an early epoch the energy    
density is dominated by the relativistic (R) particles present and the    
universe is said to be ``radiation-dominated'' (RD).  For sufficiently    
early times, when the temperature is a few times higher than the electron   
rest mass energy, these are photons, \epm pairs and, for the standard model   
of particle physics, three flavors of left-handed (\ie one helicity state)    
neutrinos (and their right-handed, antineutrinos). The hot, dense, early  
Universe is a hostile environment for complex nuclei.  At sufficiently high  
temperatures ($T \ga 80$~keV), when all the nucleons (baryons) were either  
neutrons or protons, their relative abundance was regulated by the weak  
interactions ($p + e^{-} \longleftrightarrow n + \nu_{e}, ~n + e^{+}  
\longleftrightarrow p + \bar{\nu}_{e}, ~n \longleftrightarrow p + e^{-}  
+ \bar{\nu}_{e}$); the higher mass of the neutron ensures that protons  
dominate (in the absence -- for now -- of a chemical potential for the  
electron-type neutrinos $\nu_{e}$). Below $\sim$~80~keV, the Universe has  
cooled sufficiently that the cosmic nuclear reactor can begin in earnest,  
building the lightest nuclides D, \3he, \4he, and \7li. D and \3he (also  
$^3$H) are burned very rapidly to \4he, the most tightly bound of the  
light nuclides.  The absence of a stable mass-5 nuclide ensures that in  
the expanding, cooling, early Universe, the abundances of heavier nuclides  
are greatly suppressed.  By the time the temperature has dropped below  
$\sim 30$~keV, a time comparable to the neutron lifetime, the average  
thermal energy of the nuclides and nucleons is too small to overcome  
the coulomb barriers, any remaining free neutrons decay, and BBN ceases.     
  
Among the light nuclides synthesized during BBN the relic abundances of 
D, \3he, and \7li are {\it rate limited}, determined by the competition 
between the reaction rates (which depend on the nucleon density) and the 
universal expansion rate.  Any of these nuclides are potential baryometers. 
In the expanding Universe the number densities of all particles decrease  
with time, so that the magnitude of the baryon density (or that of any other  
particle) has no meaning without also specifying {\it when it is measured}.   
To quantify the universal abundance of baryons, it is best to compare the 
baryon number density $n_{\rm B}$ to the CBR photon number density  
$n_{\gamma}$.  After \epm pairs have annihilated the ratio of these number  
densities remains effectively constant as the Universe evolves.  This ratio  
$\eta \equiv n_{\rm B}/n_{\gamma}$ is very small, so that it is convenient  
to define a quantity of order unity, 
\be 
\eta_{10} \equiv 10^{10}(n_{\rm B}/n_{\gamma}) = 274\,\Omega_{\rm B}h^{2}  
= 274\,\omega_{\rm B}, 
\label{eq:eta10} 
\ee 
where \omb is the ratio (at present) of the baryon density to the critical 
density, $h$ is the present value of the Hubble parameter in units of 100 
km s$^{-1}~$Mpc$^{-1}$, and $\omega_{\rm B} \equiv \Omega_{\rm B}h^{2}$.   
For SBBN, there is only one adjustable parameter, $\eta$ (or $\omega_{\rm B}$). 
 
As the post-\epm annihilation Universe evolves, the ratio of nucleons  
(baryons) to photons is accurately preserved so that $\eta$ at the time  
of BBN should be equal to its value at recombination (probed by the CBR)  
as well as its value today. Testing this over ten orders of magnitude  
in redshift, over a timespan of some 10 billion years, can provide a  
confirmation of the standard models of particle physics and of cosmology.     
 
\subsection{Deuterium -- The BBN Baryometer of Choice}\label{sec:deut} 
 
Deuterium is currently the baryometer of choice.  As the Universe evolves, 
galaxies form, and gas is cycled through stars.  As a consequence of its  
very weak binding, deuterium is burned at sufficiently low temperatures  
that it is completely destroyed in any gas which passes through stars.  
Therefore, in the post-BBN Universe the D abundance never exceeds its  
BBN value.  Because the evolution of D is simple (no post-BBN sources),  
decreasing since BBN, observations of D anywhere, anytime, yield a {\it  
lower} bound to its primordial abundance.  Observations of D in ``young''  
systems (high redshift or low metallicity), should reveal a deuterium  
``plateau'' at its primordial abundance.  In addition to its simple  
post-BBN evolution, deuterium is the baryometer of choice also because its 
predicted primordial abundance ($y_{\rm D} \equiv 10^{5}(D/H)_{\rm P}$) 
is sensitive to the baryon density, $y_{\rm D} \propto \eta^{-1.6}$; as  
a result, a $\sim 10\%$ measurement of \yd will lead to a $\sim 6\%$  
determination of $\eta$ (or $\Omega_{\rm B}h^{2}$). 
 
\begin{figure} 
\centering 
 \epsfysize=4.0truein 
  \epsfbox{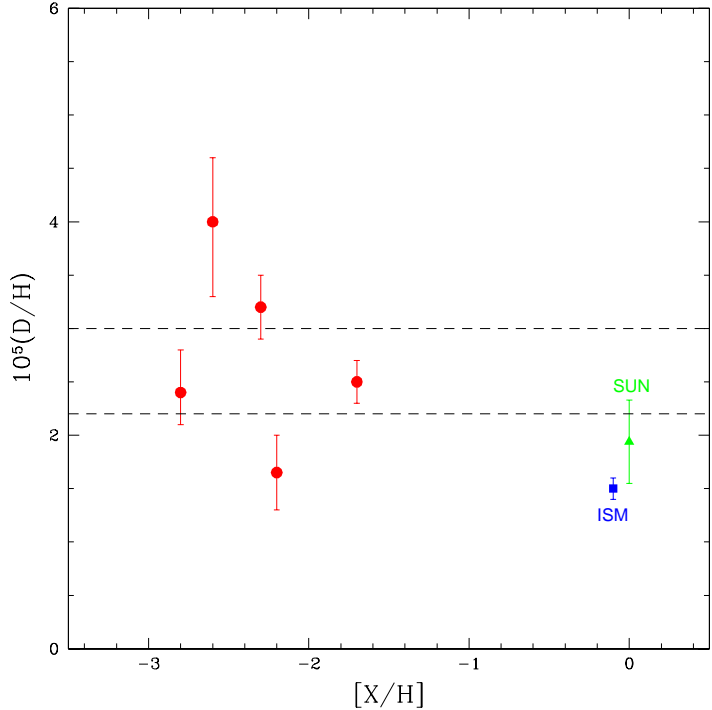} 
\vspace*{-0.15cm}   
\caption{The deuterium abundance by number with respect to hydrogen  
versus the metallicity (relative to solar on a log scale), from  
observations (as of early 2003) of QSOALS (filled circles).  Also  
shown for comparison are the D abundances for the local ISM (filled  
square) and the solar system (``Sun''; filled triangle).  The dashed  
horizontal lines represent the range of the $\pm 1\sigma$ estimate for  
the primordial deuterium abundance (\yd = $2.6 \pm 0.4$) based on the 
QSOALS data.   
\label{fig:dvssi}}   
\end{figure}

In pursuit of the most nearly primordial value of the D abundance, it  
is best to concentrate on those systems which have experienced the  
least stellar evolution.  Thus, while observations of D in the solar  
system and/or the local ISM provide useful {\it lower} bounds to the  
primordial D abundance, it is the observations of deuterium in a few  
high redshift, low metallicity, QSO absorption-line systems (QSOALS)  
which provide the most useful data.  Presently there are only five  
QSOALS with reasonably firm deuterium abundance determinations~\cite{deut}; 
the derived abundances of D are shown in Figure~\ref{fig:dvssi} along  
with the corresponding solar system and ISM D abundances. 
 
As is clear from Figure~\ref{fig:dvssi}, there is significant dispersion  
among the derived D abundances at low metallicity.  The QSOALS data fail  
to reveal the anticipated deuterium plateau, suggesting that systematic  
errors may be present, contaminating some of the determinations of the  
\di and/or \hi column densities.  Since the \di and \hi absorption spectra  
are identical, except for the wavelength/velocity offset resulting from  
the heavier reduced mass of the deuterium atom, an \hi ``interloper'',  a 
low-column density cloud shifted by $\sim 81$~km s$^{-1}$ with respect to 
the main absorbing cloud, would masquerade as D\thinspace$\scriptstyle{\rm 
I}$.  If this is not accounted for, a D/H ratio which is too high would be 
inferred.  Since there are more low-column density absorbers than those with 
high \hi column densities, absorption-line systems with somewhat lower \hi 
column density (\eg Lyman-limit systems) are more susceptible to this 
contamination than are the higher \hi column density absorbers  (\eg damped 
Ly$\alpha$ absorbers).  It is intriguing that the two QSOALS with the 
highest D/H have the lowest \hi column densities.  In contrast, for the  
damped Ly$\alpha$ absorbers an accurate determination of the \hi column  
density requires an accurate placement of the continuum, which could be  
compromised by interlopers.  This might lead to an overestimate of the  
\hi column density and a concomitant underestimate of D/H (J. Linsky,  
private communication). Again, it is intriguing that the lowest D/H  
is found for the system with the highest \hi column density.  Although  
these hints are suggestive, there is no observational evidence to support  
the exclusion of any of the current deuterium abundance determinations.   
Following the approach of O'Meara \etal~\cite{deut} and Kirkman  
\etal~\cite{deut}, the weighted mean of the D abundances for the five 
lines of sight~\cite{deut} is adopted for the primordial abundance,   
while the dispersion in the data is used to set the error: \yd = 
$2.6 \pm 0.4$.  For the same data Kirkman \etal~\cite{deut} derive 
a slightly higher mean D abundance, \yd = 2.74, by first finding the 
mean of log(y$_{\rm D}$) and then using it to compute the mean D 
abundance (\yd $ \equiv 10^{\langle \log(y_{\rm D})\rangle}$).  
 
\subsection{SBBN Baryon Density}\label{sec:bbneta} 
 
For SBBN, the baryon density parameter corresponding to the primordial  
D abundance adopted here ($y_{\rm D} = 2.6 \pm 0.4$) is $\eta_{10} =  
6.10^{+0.67}_{-0.52}$ (\obh = $\omega_{\rm B} = 0.0223^{+0.0024}_{-0.0019}$).   
This is in outstanding agreement with the Spergel \etal~estimate \cite{sperg}
of \obh $ = 0.0224 \pm 0.0009$ ($\eta_{10} = 6.14 \pm 0.25$), based largely 
on the new CBR ({\it WMAP}) data (Bennett \etal~\cite{wmap}).  The concordance 
between SBBN and the CBR is spectacular.   
 
In contrast to D, the post-BBN evolution of \3he is complex.  \3he is  
destroyed in the hotter interiors of all but the least massive stars, while  
it survives in the cooler, outer layers of nearly all stars.  Complicating 
the evolution of \3he, hydrogen burning in low mass stars leads to the  
synthesis of significant amounts of {\it new} \3he \cite{new3he}.  It is 
necessary to account for all of these effects quantitatively in the material  
returned by stars to the interstellar medium (ISM) if current Galactic 
data are to be used to infer the primordial \3he abundance.  In fact, 
current data~\cite{3he} suggest that a very delicate balance exists  
between net production and net destruction of \3he in the course of the  
evolution of the Galaxy, rendering difficult a precise estimate of the 
primordial abundance of \3he.  However, for the baryon density determined  
by D (or, the CBR), the SBBN-predicted abundance of \3he is $y_{3} = 1.0 
\pm 0.1$, which may be compared to the outer-Galaxy abundance of $y_{3} 
= 1.1 \pm 0.1$, suggested by Bania, Rood, \& Balser \cite{3he} to be 
nearly primordial.   This agreement between SBBN D and \3he is excellent.   
 
A similar scenario to that for \3he may be sketched for \7li.  As a weakly  
bound nuclide, it is easily destroyed when cycled through stars.  The high  
lithium abundances observed in the ``super-lithium-rich red giants'' reveal  
that at least some stars can produce post-BBN lithium, but a key unsolved  
issue is how much of this newly-synthesized lithium is actually returned  
to the ISM.  Furthermore, observations of nearly primordial lithium are  
limited to the oldest, most metal-poor stars in the Galaxy, stars that  
have had the most time to mix and destroy or dilute their surface lithium.   
To add to the uncertainties, errors in equivalent width measurements, in 
the temperature scales for the coolest Population II stars  and, in their 
model atmospheres contribute to the overall error budget, accounting for 
at least some of the variations among the values of the primordial abundance 
of lithium inferred by different observers.  For example, Ryan \etal~(2000) 
\cite{li} find [Li]$_{\rm P} \equiv 12 + $log(Li/H) = 2.1, while Bonifacio 
\& Molaro (1997) and Bonifacio, Molaro, \& Pasquini (1997) \cite{li} derive 
[Li]$_{\rm P} = 2.2$, and Thorburn (1994) and Bonifacio \etal~(2002) \cite
{li} infer [Li]$_{\rm P} = 2.3$.  On the basis of these not entirely 
consistent estimates, the data suggest that [Li]$_{\rm P} \approx 2.2 \pm 
0.1$.  However, for the baryon density fixed by SBBN the predicted relic 
abundance of lithium should be [Li]$_{\rm P} \approx 2.65^{+0.09}_{-0.11}$.  
Compared to expectation, the observed lithium abundance is low.  Either 
this is a challenge to SBBN or, more likely (see Pinsonneault \etal~2002 
and further references therein \cite{pinsono}), a hint about nonstandard 
stellar astrophysics.   At present the most promising approach is to use 
the observed and SBBN-predicted lithium abundances to learn about stellar 
evolution, rather than to use stellar observations to constrain the 
SBBN-inferred baryon density. 
 
\subsection{SBBN and \4he}\label{sec:4he} 
 
Unique among the relic nuclides, the primordial abundance (mass fraction) of 
\4he is not rate limited, so that \Yp provides only a weak measure of $\eta$.   
The net effect on \4he of post-BBN evolution was to burn hydrogen to helium,  
increasing the \4he mass fraction above its primordial value (Y $>$ Y$_{\rm 
P}$).  Although there are solar system and ISM (Galactic \hii regions) data 
for \4he, the key to the \Yp determinations are the observations of \4he in 
nearly unevolved, metal-poor regions.  To date these are limited to the 
observations of helium and hydrogen recombination lines in extragalactic 
\hii regions \cite{hii}.  Unfortunately, the present \hii region \4he data 
lead to a very large dispersion in the derived \Yp values, ranging from 
\Yp = 0.234$\pm 0.003$ to \Yp = 0.244$\pm 0.002$, suggesting that current 
estimates may not be limited by statistics, but by uncorrected systematic 
errors~\cite{vgs}.  Here, the compromise proposed by Olive, Steigman, and 
Walker~\cite{osw} is adopted: \Yp = 0.238$\pm 0.005$; for further discussion 
and references, see Steigman (2003a,b,c)~\cite{gs03}. 
 
For \4he there is tension between the data and SBBN.  Given the very high  
accuracy of the SBBN-abundance prediction and the very slow variation of  
\Yp with $\eta$, the SBBN-predicted primordial abundance is tightly  
constrained: Y$_{\rm P}^{\rm SBBN} = 0.248 \pm 0.001$.  Agreement with the 
adopted value of Y$_{\rm P}^{\rm OSW} = 0.238 \pm 0.005$ (or, with the 
Izotov \& Thuan \cite{hii} value of Y$_{\rm P}^{\rm IT} = 0.244 \pm 0.002$) 
is only at the $\sim 2\sigma$ level. This apparent challenge to SBBN provides 
an opportunity. 
   
Recall that once BBN begins in earnest, essentially all available neutrons  
are incorporated into \4he: \Yp is {\it neutron limited}.  The neutron to 
proton ratio when BBN commences is determined by the competition between 
the weak interaction rates and the universal expansion rate, the latter 
of which is fixed in SBBN by the standard model particle content and the 
Friedman equation.  However, if the early-Universe expansion rate differed 
from the standard model prediction, \4he provides the ideal BBN chronometer
with which to probe it.  A slower expansion leaves fewer neutrons available 
to build \4he, reducing the predicted value of Y$_{\rm P}$.  
 
The neutron-proton ratio at BBN can also be modified from its standard  
value by an {\it asymmetry} between the number densities of $\nu_{e}$ 
and $\bar{\nu}_{e}$ (``neutrino degeneracy''), described by  a chemical 
potential $\mu_{e}$ (or, equivalently, by the dimensionless degeneracy 
parameter $\xi_{e} \equiv \mu_{e}/T$).  For a {\it significant}, positive 
chemical potential ($\xi_{e} \ga 0.01$; more $\nu_{e}$ than $\bar{\nu}_{e}$) 
there are fewer neutrons than for the ``standard'' case (SBBN), leading 
to the formation of less \4he, reducing Y$_{\rm P}$.   
 
It is clear that if \4he is paired with D (the BBN baryometer), together 
they can constrain cosmological models with nonstandard expansion rates 
(or particle content) or neutrino degeneracy.  After reviewing the BBN  
and CBR constraints on models with nonstandard, early-Universe expansion  
rates (see Steigman 2003a,b,c \cite{gs03} and further references therein),  
this review concentrates on using BBN and the CBR to constrain any 
neutrino degeneracy (see Barger \etal~2003a \cite{barger03a} for further 
details and references to related work). 
  
\section{Early Universe Expansion Rate}\label{sec:S} 
 
The expansion rate (measured by the Hubble parameter $H$) is related to  
the energy density through the Friedman equation ($H^{2}={8\pi G \over  
3}\rho$).  During BBN the Universe is ``radiation-dominated'' (RD);  
the energy density ($\rho_{\rm R}$) dominated by the relativistic  
particles present. For the standard model (SBBN) there are three families  
of light (relativistic) neutrinos (N$_{\nu} = 3$).  For models with  
nonstandard, early-Universe expansion rates it is convenient to  
introduce the {\it expansion rate factor} $S \equiv H'/H = t/t'$.   
One possible origin for a nonstandard expansion rate ($S \neq 1$) is  
a modification of the energy density by the presence of ``extra''  
relativistic particles $X$: $\rho_{\rm R} \rightarrow \rho_{\rm R}'  
= \rho_{\rm R} + \rho_{X}$.  If the additional energy density is  
normalized to that which would be contributed by additional flavors  
of (decoupled) neutrinos (Steigman, Schramm, \& Gunn~\cite{ssg}),  
$\rho_{X} \equiv \Delta N_{\nu}\rho_{\nu}$, and $N_{\nu} = 3 +  
\Delta N_{\nu}$. It must be emphasized that the fundamental physical  
parameter is $S$, the expansion rate factor, which may be related to 
\Deln ({\it nonlinearly}) by 
\be 
S_{pre} \equiv (H'/H)_{pre} = (1 + 0.163\Delta N_{\nu})^{1/2}\,; \, \, \,  
S_{post} \equiv (H'/H)_{post} = (1 + 0.135\Delta N_{\nu})^{1/2}, 
\label{eq:sx} 
\ee 
where the subscripts ``$pre$'' and ``$post$'' reflect the values prior  
to, and after \epm annihilation, respectively.  However, any term in a  
modified Friedman equation which scales like radiation (decreases as  
the fourth power of the scale factor), such as may be due to higher  
dimensional effects as in the Randall-Sundrum~\cite{rs} model, will 
change the standard-model expansion rate ($S \neq 1$) and may be related 
to an {\it equivalent} \Deln (which could be {\it negative} as well as 
positive; $S > 1$ and $S < 1$ are both possible) through Eq.~\ref{eq:sx}. 
 
\subsection{BBN Constraints on $S$}\label{sec:bbns} 
  
\begin{figure}   
\vspace*{-1.25cm}   
\begin{center} 
\epsfysize=4.0truein 
  \epsfbox{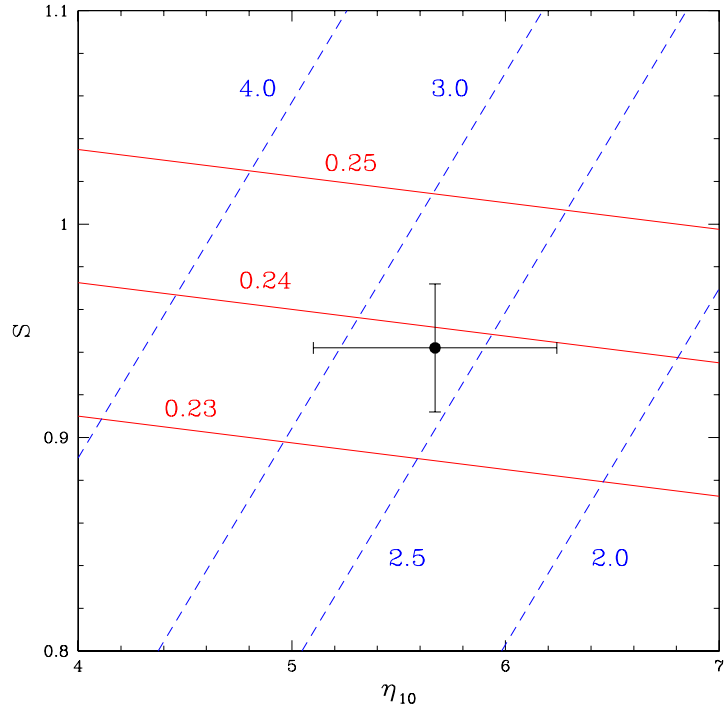} 
\end{center} 
\vspace*{-0.7cm}   
\caption{Isoabundance curves for D and \4he in the expansion rate 
parameter ($S$) -- baryon abundance parameter ($\eta_{10}$) plane.  
The dashed curves are for the deuterium abundances; the numbers are  
the values of \yd $\equiv 10^{5}$(D/H).  The solid curves are for 
the helium-4 mass fractions; the numbers correspond to Y$_{\rm P}$. 
The filled circle with error bars is for the adopted D and \4he 
abundances (see the text).  
} 
\label{fig:isosd}  
\end{figure}  
 
\begin{figure}   
\vspace*{0.3cm}   
\begin{center} 
\epsfysize=4.0truein 
  \epsfbox{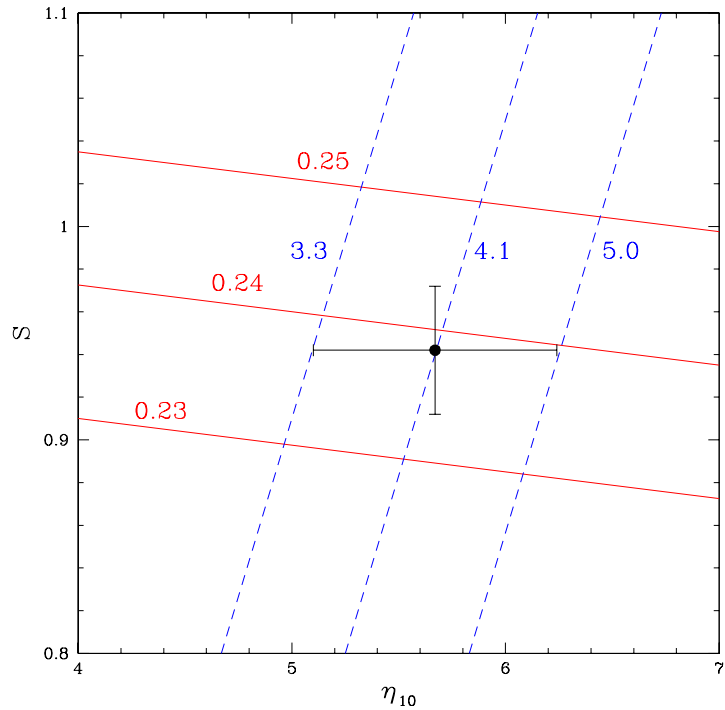} 
\end{center} 
\vspace*{-0.7cm}   
\caption{Isoabundance curves for \7li and \4he in the expansion rate 
parameter ($S$) -- baryon abundance parameter ($\eta_{10}$) plane.  
The dashed curves are for the lithium abundances; the numbers are  
the values of \y7 $\equiv 10^{10}$(Li/H).  The solid curves are for 
the helium-4 mass fractions; the numbers correspond to Y$_{\rm P}$. 
The filled circle with error bars is for the adopted D and \4he 
abundances (see the text).  
} 
\label{fig:isosli}  
\end{figure}  
 
The qualitative effects of a nonstandard expansion rate on the relic 
abundances of the light nuclides may be understood as follows.  For  
the baryon abundance range of interest ($1 \la \eta_{10} \la 10$)  
the relic abundances of D and \3he are decreasing functions of $\eta$,  
revealing that in this range D and \3he are being destroyed.  A faster  
than standard expansion ($S > 1$) leaves less time for destruction, so  
more D and \3he survive.  At the higher values of $\eta$ suggested by  
D, the \7li abundance increases with $\eta$, so that less time available  
($S > 1$) results in less production and a {\it smaller} \7li relic 
abundance. Generally, these effects on the relic abundances of D, \3he,  
and \7li are subdominant to their dependences on the baryon density.   
Not so for \4he, whose relic abundance is weakly (logarithmically)  
dependent on the baryon density, while being strongly affected by the  
early-Universe expansion rate.  A faster expansion leaves more neutrons  
available to be incorporated into \4he; to a good approximation,  
$\Delta$Y $\approx 0.16\,(S-1)$.  Using D as our baryometer and \4he  
as our chronometer, the BBN constraints on $\eta$ and $S$ are shown 
in the $S - \eta$ plane in Figure \ref{fig:isosd}, employing 
approximations to the D and \4he isoabundance contours \cite{ks03} 
which are typically good to a few percent or better. 
 
It is clear from Figure \ref{fig:isosd} that the adopted D and \4he 
abundances favor a slower-than-standard expansion rate.  The best 
fit point is at $S = 0.94$ ($\Delta$N$_{\nu} = -0.7$).  However, 
the $2\sigma$ range in \Nnu extends from \Nnu = 1.7 to \Nnu = 3.0, 
consistent with the standard case of \Nnu = 3. 
 
In Figure \ref{fig:isosli} are shown the corresponding \7li -- \4he 
isoabundance contours in the $S - \eta$ plane, along with the best  
fit point and $1\sigma$ error from the adopted D and \4he abundances. 
The constraints on $S$ and $\eta$ derived from D and \4he lead to 
a predicted primordial lithium abundance in the range $2.5 \la [{\rm 
Li}]_{\rm P} \la 2.7$, not very different from the range for SBBN.  
It is clear that BBN with a nonstandard expansion rate, consistent 
with D and \4he, cannot relieve the tension between the BBN-predicted 
\7li abundance and that inferred from metal-poor stars in the Galaxy.   
As mentioned earlier (\S \ref{sec:bbneta}), the solution likely  
lies with the stellar astrophysics. 
 
\subsection{Joint BBN and CBR Constraints on $S$ and $\omega_{\rm B}$} 
 
\begin{figure}[t]   
\centering 
\vspace*{0.2cm} 
 \epsfysize=3.1truein 
  \epsfbox{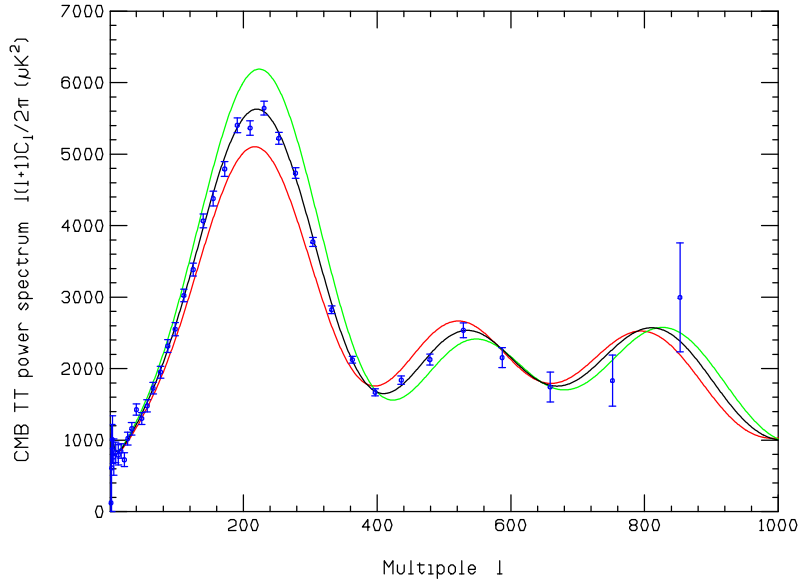} 
\vspace*{-0.2cm} 
\caption{The CBR temperature fluctuation anisotropy spectra for three  
choices of the baryon density parameter $\omega_{\rm B} = 0.018$, 0.023,  
0.028, in order of increasing height of the first peak. The WMAP data 
points \cite{wmap} are shown. 
\label{fig:etaspectrum}}    
\end{figure} 
 
\begin{figure}[t]   
\centering 
\vspace*{-0.8cm} 
 \epsfysize=3.1truein 
  \epsfbox{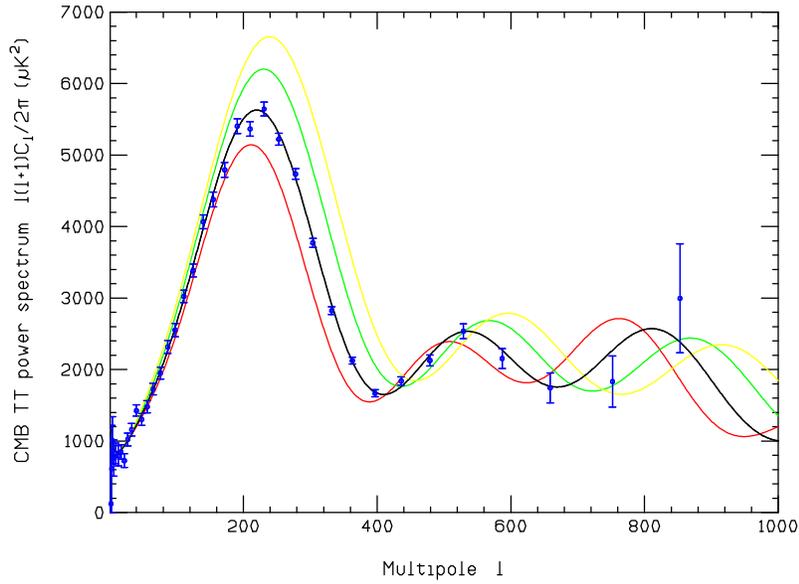} 
\vspace*{-0.2cm} 
\caption{The CBR temperature fluctuation anisotropy spectra for four  
choices of N$_{\nu}$ = 1, 2.75, 5, 7, in order of increasing height  
of the first peak. The WMAP data points \cite{wmap} are shown. 
\label{fig:nnuspectrum}}    
\end{figure}

\begin{figure}[t]   
\centering 
\vspace*{0.09cm} 
 \epsfysize=3.05truein 
  \epsfbox{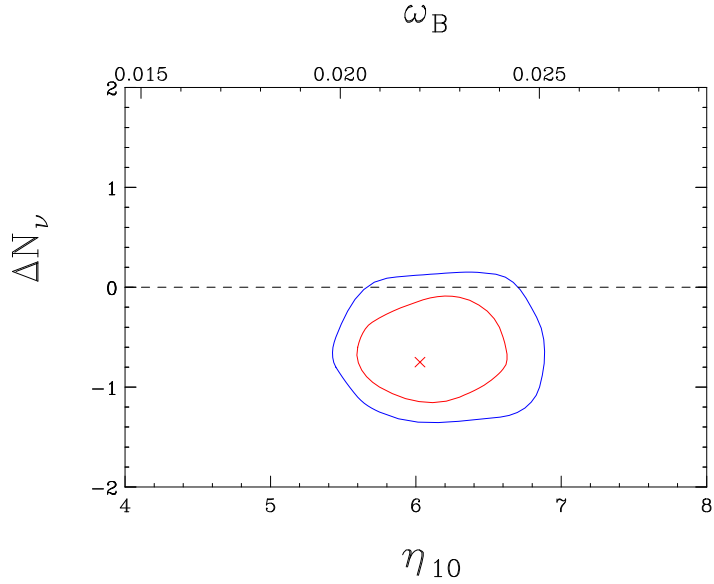} 
\vspace*{-0.1cm} 
\caption{The 1$\sigma$ and 2$\sigma$ contours in the $\eta$ ($\omega_{\rm B}$)  
-- \Deln plane for the joint BBN -- CBR ({\it WMAP}) fit.  
\label{fig:jointcontours}}    
\end{figure}

The CBR temperature anisotropy and polarization spectra are sensitive  
to the baryon density and to the early-Universe (RD) expansion rate (see,  
\eg Barger \etal~(2003b) \cite{barger03b}, and references to related  
work therein).  Increasing the baryon density increases the inertia of  
the baryon-photon fluid, shifting the positions and the relative sizes  
of the odd and even acoutic peaks; see Figure \ref{fig:etaspectrum}.   
Changing the expansion rate (or, the relativistic particle content)  
changes the redshift of matter-radiation equality, also shifting the 
locations of the peaks.  Since any change in \Deln (or $S$) can be 
mimicked by a corresponding change in the total matter density $w_{\rm M} 
\equiv \Omega_{\rm M}h^{2}$, there is a degeneracy in the CBR anisotropy  
spectrum between $S$ and $w_{\rm M}$ which can be broken using the HST 
Key Project determination of the Hubble parameter \cite{barger03b}.  
CBR temperature anisotropy spectra for four choices of \Nnu are shown 
in Figure \ref{fig:nnuspectrum}.  Although the CBR temperature anisotropy  
spectrum is a less sensitive early-Universe chronometer than is BBN  
(\4he), the WMAP data may be used to identify allowed regions in the  
\Deln (or $S$) -- $\eta$ plane, similar to those from BBN using D and  
\4he.  There is excellent overlap between the $\eta$ -- \Deln ($S$)  
likelihood contours from BBN and those from the CBR (see Barger  
\etal~(2003b) \cite{barger03b}); this variant of SBBN ($S \neq 1$)  
is consistent with the CBR.  In Figure \ref{fig:jointcontours} (from  
Barger \etal~(2003b) \cite{barger03b}) the confidence contours in the  
$\eta$ -- \Deln plane are shown for a joint BBN -- CBR fit.  Although  
the best fit value for \Deln is negative (driven largely by the adopted  
value for Y$_{\rm P}$), the standard-model value of \Deln = 0 ($S = 
1$) is quite acceptable.

\section{Neutrino Asymmetry}\label{sec:degen} 
 
Oscillations among the three known (active) neutrinos ($\nu_{e}$,  
$\nu_{\mu}$, $\nu_{\tau}$) will equilibrate any pre-existing  
asymmetries in any (all) of them to the level of the electron  
neutrino degeneracy prior to BBN~\cite{equilibrate}.  Thus,  
the magnitude of the electron neutrino degeneracy allowed by  
BBN (and the CBR) is of special interest to any determination  
of constraints on the universal lepton asymmetry.  As will be  
seen below (see Barger \etal~(2003a) \cite{barger03a}), the  
BBN constraints provide the limit $\xi_{e} \approx \xi_{\mu} 
\approx \xi_{\tau} \la 0.3$.  This constraint is important because  
{\it any} neutrino degeneracy ($\xi_{e} >$ or $< 0$) implies  
a higher energy density in the relic neutrinos compared to the  
standard, $\xi_{e} = 0$ (or, $|\xi_{e}| \ll 1$) case; for the  
three active neutrinos with $\xi_{e} = \xi_{\mu} = \xi_{\tau}  
\la 0.3$, the extra energy, expressed in terms of an equivalent  
number of extra neutrinos, is limited to \Deln $\approx {90  
\over 7}({\xi_{e} \over \pi})^{2} \la 0.1$.  As a result, the  
effect of neutrino degeneracy on the CBR, limited to its effect  
on $S$, is quite small.  Thus, while BBN and the CBR are both 
sensitive to the baryon density and the expansion rate, only BBN
is sensitive to the range of neutrino degeneracy of interest.  
Therefore, the CBR can provide independent constraints on $S$ 
and $\eta$, allowing BBN to be used to limit neutrino degeneracy. 
 
\subsection{BBN and $\xi_{e}$} 
 
\begin{figure}   
\vspace*{-1.1cm}   
\begin{center} 
\epsfysize=4.0truein 
  \epsfbox{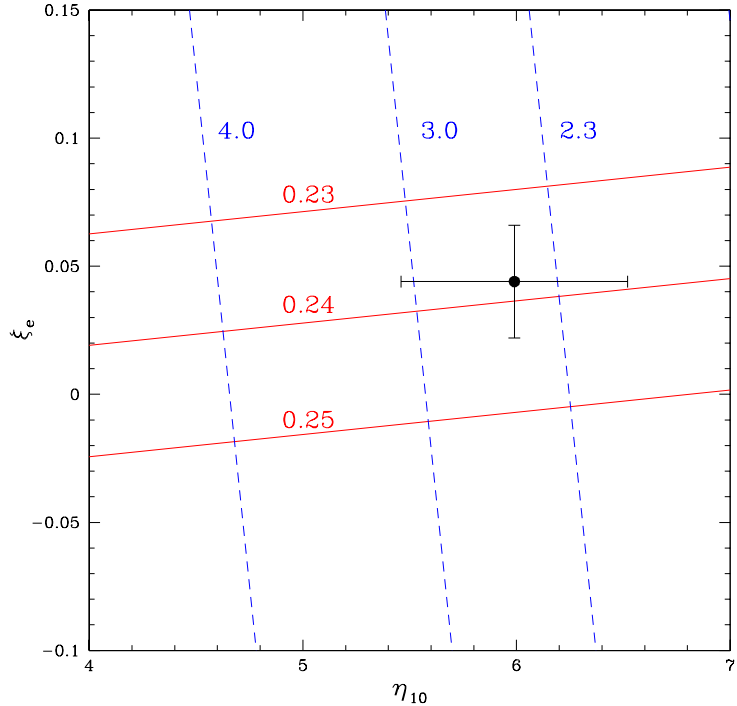} 
\end{center} 
\vspace*{-0.7cm}   
\caption{Isoabundance curves for D and \4he in the neutrino asymmetry  
parameter ($\xi_{e}$) -- baryon abundance parameter ($\eta_{10}$) plane.  
As in Fig.~\ref{fig:isosd}, the dashed curves are for the deuterium  
abundances and the solid curves are for the helium-4 mass fractions. The  
filled circle with error bars is for the adopted D and \4he abundances.  
} 
\label{fig:isoxid}  
\end{figure}  
 
\begin{figure}   
\vspace*{-1.1cm}   
\begin{center} 
\epsfysize=4.0truein 
  \epsfbox{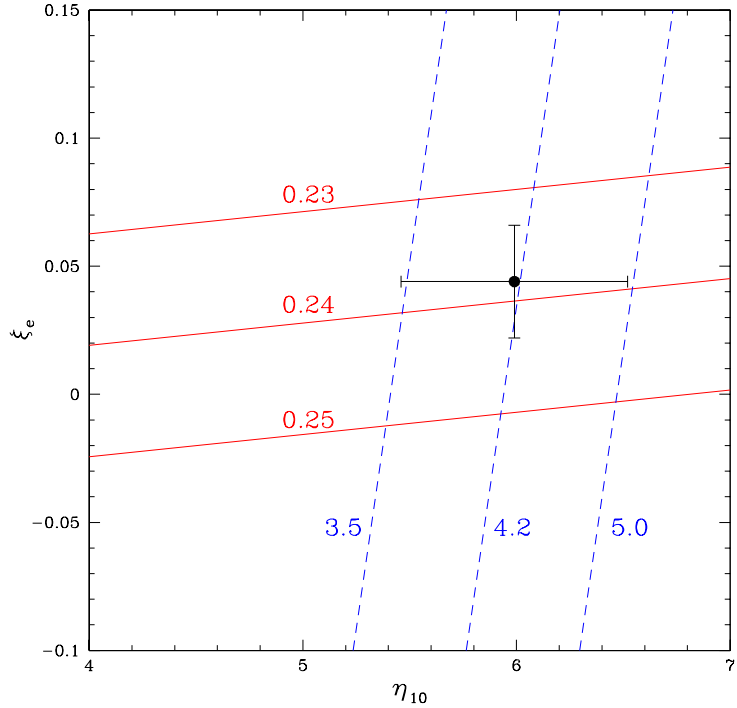} 
\end{center} 
\vspace*{-0.7cm}   
\caption{Isoabundance curves for \7li and \4he in the neutrino asymmetry  
parameter ($\xi_{e}$) -- baryon abundance parameter ($\eta_{10}$) plane.  
As in Fig.~\ref{fig:isosli}, the dashed curves are for the lithium  
abundances and the solid curves are for the helium-4 mass fractions. The  
filled circle with error bars is for the adopted D and \4he abundances.  
} 
\label{fig:isoxili}  
\end{figure}

As discussed in \S \ref{sec:4he}, \Yp is sensitive to any neutrino  
asymmetry.  More $\nu_{e}$ than $\bar{\nu}_{e}$ drives down  
the neutron-to-proton ratio, leaving fewer neutrons available to build  
\4he; to a good approximation $\Delta$Y $\approx -0.23\,\xi_{e}$  
\cite{ks}.  In contrast, the relic abundances of D, \3he, and  
\7li are very insensitive to $\xi_{e} \neq 0$, so that when paired with  
\4he, they can simultaneously constrain the baryon density and the  
electron-neutrino asymmetry. In analogy with Figures \ref{fig:isosd} 
and \ref{fig:isosli} for $S$ versus $\eta$, in Figures \ref{fig:isoxid}  
and \ref{fig:isoxili} are shown the approximate D -- \4he and \7li -- \4he  
isoabundance curves \cite{ks03} in the $\xi_{e} - \eta_{10}$ plane for  
the case where the expansion rate is fixed at its standard value, $S = 1$.  
  
While the best fit point is at a non-zero neutrino aysmmetry, $\xi_{e}  
= 0.044$ (and $\eta_{10} = 6.0$), $\xi_{e} = 0$ is consistent at the 
$\sim 2\sigma$ level (and $\xi_{e} \ga 0.09$ is excluded at $\sim 2\sigma$). 
It is clear from Figure \ref{fig:isoxili} that the combination of neutrino 
asymmetry and baryon density which reconciles D and \4he, leaves the 
BBN-predicted \7li abundance virtually unchanged from its SBBN value; 
the BBN-predicted \7li abundance is still too large compared to that 
inferred from the metal-poor halo stars in the Galaxy. 
 
\section{Joint BBN and CBR Constraints on $\xi_{e}$}  
 
Now consider the case where {\it both} $\xi_{e}$ and \Deln ($S$) are free 
to vary.  As \Nnu increases, so too will the best fit values of $\eta$ 
and $\xi_{e}$.  If \Nnu increases, the early universe expands more rapidly, 
leaving {\it less} time to burn D.  As a result, for a {\it fixed} baryon 
density parameter \yd {\it increases}.  To reduce \yd back to its observed  
value, the baryon density parameter must increase.  But, the combination of 
an increased baryon-to-photon ratio and \Nnu $> 3$, raises the predicted 
primordial abundance of \4he, requiring a larger $\xi_{e}$ to reconcile the 
BBN predictions with the data.  For {\it any} (reasonable) choice of \Nnu  
there is always a pair of $\{\eta_{10},\xi_{e}\}$ values for which {\it 
perfect} agreement with the observed D and \4he abundances can be obtained.   
It is, of course, not surprising that with three parameters and two  
constraints, such a fit can be found.  Another observational constraint  
is needed to break this degeneracy and to simultaneously constrain $\eta$, 
$S$, and $\xi_{e}$.  While neither \3he nor \7li can provide the needed 
constraint, the CBR temperature anisotropy spectrum, which is sensitive 
to $\eta$ and $S$ but not to $\xi_{e}$, can (see Barger \etal~2003a 
\cite{barger03a}).   
  
\begin{figure} 
\begin{center} 
\epsfxsize=4.5in 
\epsfbox{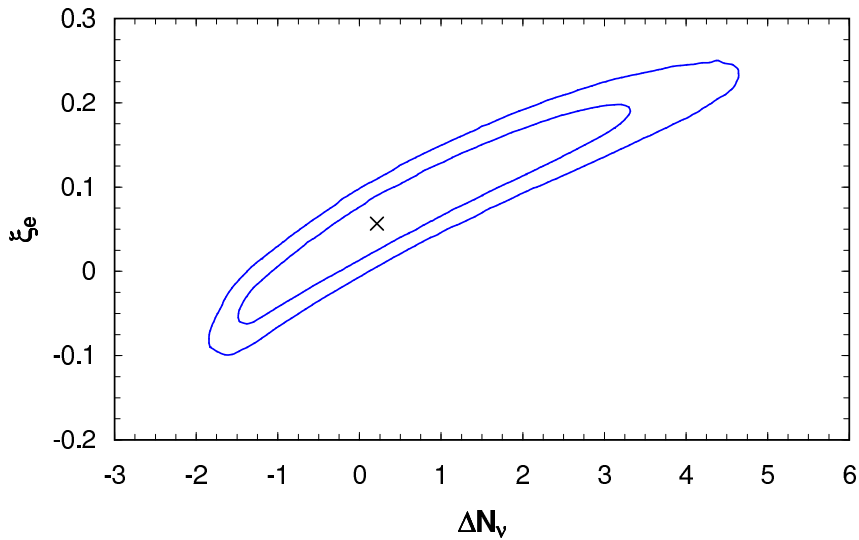} 
\end{center} 
\vspace*{-0.6cm}   
 \caption{The $1\sigma$ and $2\sigma$ contours in the 
\Deln -- $\xi_{e}$ plane using WMAP data and BBN with  
the adopted D and \4he abundances.  The best-fit point  
is marked with a cross. 
}                     
\label{fig:xivsnnu} 
\end{figure} 
 
With the primordial abundances of D and \4he as fixed constraints, 
increased values of $\xi_{e}$ may be compensated by increased values 
of \Deln and $\eta$.  This degeneracy is broken by the WMAP CBR data 
which provides restrictions on the relation between \Deln and $\eta$. 
The expanded ranges of $\xi_{e}$ and \Deln which are consistent with 
both the CBR and BBN are shown in Figure \ref{fig:xivsnnu}.  Values 
of \Deln in the range $-2 \la \Delta N_{\nu} \la 5$ are allowed,
{\it provided that} the neutrino asymmetry parameter lies in the range 
$-0.1 \la \xi_{e} \la 0.3$ (and, vice-versa).
While permitting three free parameters ($\eta$, $\Delta$N$_{\nu}$, 
$\xi_{e}$) weakens the constraints on any one of them, the {\it 
combination} of BBN and the CBR can provide meaningful bounds.  
Allowing nonzero values of $\xi_{e}$ weakens the previous BBN 
restrictions (\S \ref{sec:bbns}) on \Deln (or, on the early-Universe 
expansion rate parameter $S$).  Nonetheless, after marginalizing 
over $\xi_{e}$, the 95\% confidence level range for \Deln is $-1.7 
\la \Delta N_{\nu} \la 4.1$ (Barger \etal~(2003a) \cite{barger03a}). 
  
\section{Conclusions}  
 
For the standard models of cosmology and particle physics, the  
SBBN-predicted primordial abundances of D, \3he, \4he, and \7li 
depend on only one free parameter, the baryon abundance parameter.  
Of the light nuclides, deuterium is the baryometer of choice; the 
SBBN-derived baryon density based on D, $\eta_{10}({\rm SBBN}) =  
6.10^{+0.67}_{-0.52}$, is in excellent (exact!) agreement with  
that derived from non-BBN, mainly CBR, data (Spergel \etal~(2003)  
\cite{sperg}): $\eta_{10}({\rm CBR}) = 6.14 \pm 0.25$.  Unique among 
the relic nuclides,  \4he is an excellent chronometer and is also 
sensitive to any $\nu_{e} - \bar{\nu}_{e}$ asymmetry.  While for SBBN 
the predicted primordial \4he mass fraction is slightly low compared 
to available data, the uncertainties in the observationally inferred 
primordial value are likely dominated by systematics.  However, if the 
tension between SBBN D and \4he persists, it could be relieved by a 
nonstandard expansion rate, by a neutrino asymmetry, or by both acting 
together.  Neither of these options can reconcile the BBN prediction 
with the low abundance of lithium inferred from observations of 
metal-poor stars, suggesting that the resolution of this conflict 
is likely to be found in the stellar astrophysics \cite{pinsono}.  
Taken together, the CBR (WMAP) and BBN (D \& \4he) are consistent 
with $\eta_{10} \approx 6$ ($\omega_{\rm B} \approx 0.022$) and 
(for $\xi_{e} \equiv 0$) $1.7 \leq {\rm N}_{\nu} \leq 3.0$, or 
(for $S \equiv 1$) $0 \la \xi_{e} \la 0.09$ (both at $\sim 2\sigma$).  
For both \Deln and $\xi_{e}$ free, $1 \la {\rm N}_{\nu} \la 7$ and 
$-0.1 \la \xi_{e} \la 0.3$ are permitted by the joint BBN and CBR 
data. 
 
In the current, data-rich era of cosmological research, BBN continues  
to play an important role.  The spectacular agreement among the estimates
of the baryon density inferred from processes occurring at widely separated
epochs confirms the general features of the standard models of cosmology 
and of particle physics.
Although there have been many successes, much remains to be done.  Whether 
the resolutions of the current challenges are observational or theoretical, 
cosmological or astrophysical, the future is bright.  
  
\acknowledgements{  
I am grateful to all my collaborators, past and present, and I thank  
them for their contributions to the material reviewed here.  Many of  
the quantitative results (and figures) presented here are from recent  
collaborations or discusions with V. Barger, J.P. Kneller, D. Marfatia,  
K.A. Olive, R.J. Scherrer, and T.P. Walker.  My research is supported  
at OSU by the DOE through grant DE-FG02-91ER40690. This manuscript was  
prepared while I was visiting the Instituto Astr$\hat{\rm o}$nomico e  
Geof\' \i sico of the Universidade de S$\tilde{\rm a}$o Paulo; I thank  
them for their hospitality. 
}

\vfill  

\begin{thebibliography}{}{  
 
 
\bibitem{gs03} Steigman, G., 2003a, Big Bang Nucleosynthesis:  
 Probing The First 20 Minutes, To appear in the Carnegie  
 Observatories Astrophysics Series, Vol. 2: Measuring and Modeling  
 the Universe; ed. W. L. Freedman (Cambridge: Cambridge University  
 Press) (astro-ph/0307244); Steigman, G., 2003b, Primordial  
 Nucleosynthesis, To appear in the Proceedings of the May 2003  
 STScI Symposium, ``The Local Group As An Astrophysical Laboratory";  
 ed. M. Livio (Cambridge: Cambridge University Press) (astro-ph/0308511);  
 Steigman, G., 2003c, The Baryon Budget From BBN And The CBR, To  
 appear in the proceedings of the XV Rencontres de Blois, ``Physical  
 Cosmology: New Results in Cosmology and the Coherence of the  
 Standard Model'' (astro-ph/0309338)   
 
\bibitem{wmap} Bennett, C. L. \etal, 2003, Astrophys. J. Suppl., 
148, 1. 
  
\bibitem{sperg} Spergel, D. N. \etal, 2003, Astrophys. J. Suppl., 
148, 175.  
 
\bibitem{barger03a} Barger, V., Kneller, J. P., Marfatia, D.,  
 Langacker, P. \& Steigman, G., 2003a, Phys. Lett. B, 569, 123. 
 
\bibitem{ks} Kang, H.-S. \& Steigman, G., 1992, Nucl. Phys. B., 372, 494. 
 
\bibitem{deut} Burles, S. \& Tytler, D., 1998a, Astrophys. J., 499, 699;  
 {\em ibid}, Astrophys. J., 507, 732; O'Meara, J. M., Tytler, D., Kirkman,  
 D., Suzuki, N., Prochaska, J. X., Lubin, D., \& Wolfe, A. M., 2001,  
 Astrophys. J., 552, 718; Pettini M. \& Bowen, D. V., 2001, Astrophys.  
 J., 560, 41; Kirkman, D., Tytler, D., Suzuki, N., O'Meara, J. M., \&  
 Lubin, D., 2003, Astrophys. J. Suppl., submitted (astro-ph/0302006). 
 
\bibitem{new3he} Iben, I. I., 1967, Astrophys. J., 147, 624; Rood, R. T., 
 1972, Astrophys. J., 177, 681; Rood, R. T., Steigman, G. \& Tinsley, 
 B. M., 1976, Astrophys. J. Lett., 207, L57; Dearborn, D. S. P., Schramm, 
 D. N., \& Steigman, G., 1986, Astrophys. J., 203, 35. 
 
\bibitem{3he} Bania, T. M., Rood, R. T., \& Balser, D., 2002, Nature,  
 415, 54. 
 
\bibitem{li} Thorburn. J. A., 1994, Astrophys. J., 421, 318; Bonifacio,  
 P. \& Molaro, P., 1997, MNRAS, 285, 847; Bonifacio, P., Molaro, P.,  
 \& Pasquini, L., 1997, MNRAS, 292, L1; Ryan, S. G., Beers, T. C., Olive,  
 K. A., Fields, B. D., \& Norris, J. E., 2000, Astrophys. J. Lett., 530,  
 L57; Bonifacio, P., \etal, 2002, Astron. \& Astrophys., 390, 91. 
 
\bibitem{pinsono} Pinsonneault, M. H., Steigman, G., Walker, T. P., \&  
 Narayanan, V. K., 2002, Astrophys. J., 574, 398. 
 
\bibitem{hii} Olive, K. A. \& Steigman, G., 1995, Astrophys. J. Suppl.,  
 97, 49; Izotov, Y. I., Thuan T. X. \& Lipovetsky V. A., 1997, Astrophys.  
 J. Suppl., 108, 1; Olive, K. A., Skillman, E., \& Steigman, G., 1997,  
 Astrophys. J., 483, 788; Izotov, Y. I. \& Thuan, T. X., 1998, Astrophys.  
 J., 500, 188. 
 
\bibitem{vgs} Viegas, S. M., Gruenwald, R., \& Steigman, G., 2000,  
 Astrophys. J., 531, 813.  
 
\bibitem{osw} Olive, K. A., Steigman, G., \& Walker, T. P., 2000, Phys.  
 Rep., 333, 389. 
 
\bibitem{ssg} Steigman, G., Schramm, D. N., \& Gunn, J. E., 1977,  
 Phys. Lett. B, 66, 202. 
 
\bibitem{rs} Randall, L. \& Sundrum, R., 1998a, Phys. Rev. Lett., 83,  
 3370; {\it ibid}, 1998b, Phys. Rev. Lett., 83, 4690; Binetruy, P.,  
 Deffayet, C., Ellwanger, U., \& Langlois, D., 2000, Phys. Lett.  
 B, 477, 285; Cline, J. M., Grojean, C., \& Servant, G., 2000, Phys. Rev.  
 Lett., 83, 4245. 
 
\bibitem{ks03} Kneller, J. P. \& Steigman, G., 2003, BBN For Pedestrians, 
 In preparation. 
 
\bibitem{barger03b} Barger, V., Kneller, J. P., Lee, H.-S., Marfatia,  
 D., \&  Steigman, G., 2003b, Phys. Lett. B, 566, 8. 
 
\bibitem{equilibrate} Lunardini, C. \& Smirnov, A. Y., 2001, Phys. Rev.  
 D, 64, 073006; Dolgov, A. D., Hansen, S. H., Pastor, S., Petcov, S. T., 
 Raffelt, G. G., \& Semikoz, D. V., 2002, Nucl. Phys. B, 632, 363;  
 Abazajian, K. N., Beacom, J. F., \& Bell, N. F., 2002, Phys. Rev. D,  
 66, 013008; Wong, Y. Y. 2002, Phys. Rev. D, 66, 025015. 
}  
\end{thebibliography}
\end{document}